\documentclass[nocopyright]{cimento}

\usepackage{amsmath}
\usepackage{graphicx}
\usepackage{subcaption}
\usepackage{float}
\usepackage{cleveref}

\usepackage{comment}
\usepackage{caption}
\usepackage{xcolor}

\usepackage{tikz}

\usepackage{tikz-feynman}
\usetikzlibrary{positioning}
\usetikzlibrary{arrows.meta, decorations.pathmorphing}
\tikzfeynmanset{compat=1.1.0}

\usepackage{lineno}

\newcommand{\novbb}{$0\nu\beta\beta$}
\newcommand{\twovbb}{$2\nu\beta\beta$}
\newcommand{\tild}{~}
\crefname{subfigure}{Fig.}{Figs.}
\Crefname{subfigure}{Fig.}{Figs.}

\title{First results from LEGEND-200 on the search for neutrinoless double beta decay}
\author{G.~Saleh\from{ins:x}\from{ins:y}\from{ins:z} \thanks{E-mail \texttt{giovanna.saleh@phd.unipd.it} } on behalf of the LEGEND Collaboration}

\instlist{
	\inst{ins:x} Dipartimento di Fisica e Astronomia, Universit\`a degli Studi di Padova - Padova, Italy
	
	\inst{ins:y} INFN Sezione di Padova - Padova, Italy
	
	\inst{ins:z} Physik-Institut, Universit\"at Z\"urich - Z\"urich, Switzerland}

\begin{document}

\maketitle

\begin{abstract}
Neutrinoless double beta decay (\novbb) is a rare process which could take place if neutrinos are Majorana fermions: the observation of this decay would provide unambiguous evidence for the existence of new Physics beyond the Standard Model, as it entails a two-units lepton number violation.
The LEGEND experiment (Large Enriched Germanium Experiment for Neutrinoless $\beta\beta$ Decay) searches for the \novbb{} of $^{76}$Ge using High Purity Germanium detectors enriched in  $^{76}$Ge beyond the 86\%.
The LEGEND project foresees two phases, LEGEND-200, aiming at a final 3$\sigma$ discovery sensitivity beyond 10$^{27}$ yr, 
and LEGEND-1000, aiming at a final sensitivity beyond 10$^{28}$ yr. 
LEGEND-200 started taking data in 2023 at Laboratori Nazionali del Gran Sasso (LNGS): in about one year it collected a total physics exposure of 61 kg yr, with a background index of $0.5^{+0.3}_{-0.2} \cdot 10^{-3}$ counts/(keV kg yr) in the so-called \texttt{golden dataset} and of $1.3^{+0.8}_{-0.5} \cdot 10^{-3}$ counts/(keV kg yr) in the so-called \texttt{silver dataset}. The performed statistical analysis finds no evidence for a \novbb{} signal and sets a lower limit on its half life to $T_{1/2}^{0\nu} > 0.5 \cdot 10^{26}$ yr at 90\% CL. A combined analysis of the three germanium-based  \novbb{} experiments GERDA, MAJORANA Demonstrator and LEGEND-200 provides a lower limit of $T_{1/2}^{0\nu} > 1.9 \cdot 10^{26}$ yr at 90\% CL.

\end{abstract}

\vspace*{-0.5cm}

\section{Neutrinoless double beta decay}

\begin{figure}[H]
	\hspace*{1.7cm}
	\resizebox{0.8\textwidth}{!}{
	\begin{minipage}{1\textwidth}
	\begin{subfigure}[b]{0.25\textwidth}
		\centering
		\begin{tikzpicture}[scale=0.95]
			\begin{feynman}
				\vertex (n1) at (0,3.5) {\( \text{n} \)};
				\vertex[right=4cm of n1] (p1) {\( \text{p} \)};
				\vertex (n2) at (0,0) {\( \text{n} \)};
				\vertex[right=4cm of n2] (p2) {\( \text{p} \)};
				
				\vertex at ($(n1)!0.4!(p1)$) (mid1) {};
				\vertex at ($(n2)!0.4!(p2)$) (mid2) {};
				
				\vertex[below right=1.5cm of mid1, dot, minimum size=0pt, inner sep=0pt] (e1) {};
				\vertex[above right=0.4cm and 1.3cm of e1] (electron1) {\( e^- \)};
				\vertex[below right=0.4cm and 1.3cm of e1] (nue1) {\( \bar{\nu}_e \)};
				
				\vertex[above right=1.5cm of mid2, dot, minimum size=0pt, inner sep=0pt] (e2) {};
				\vertex[above right=0.4cm and 1.3cm of e2] (nue2) {\( \bar{\nu}_e \)};
				\vertex[below right=0.4cm and 1.3cm of e2](electron2) {\( e^- \)};
				\diagram* {
					(n1) -- [draw = none](p1),
					(n2) -- [draw = none](p2),
					
					(mid1) -- [boson, edge label'=\( W^- \)] (e1),
					(e1) -- [fermion] (electron1),
					(e1) -- [anti fermion] (nue1),
					
					(mid2) -- [boson, edge label=\( W^- \)] (e2),
					(e2) -- [fermion] (electron2),
					(e2) -- [anti fermion] (nue2),
				};
				\foreach \shift in {-0.1, 0, 0.1} {
					\draw[thick] ($(n1)+(0.2,\shift)$) -- ($(p1)+(-0.2,\shift)$);
				}
				
				\foreach \shift in {-0.1, 0, 0.1} {
					\draw[thick] ($(n2)+(0.2,\shift)$) -- ($(p2)+(-0.2,\shift)$);
				}
				\node[draw=none, fill=cyan!20, rounded corners, minimum width=1.7cm, minimum height=1cm] at (0,2.5) {\Large \( 2\nu\beta\beta \)};
			\end{feynman}
		\end{tikzpicture}
		\caption{}
		\label{2vbb}
	\end{subfigure}
	\hspace*{3.5cm}
	\begin{subfigure}[b]{0.25\textwidth}
		\centering
		\begin{tikzpicture}[scale=0.95]
			\begin{feynman}
				\vertex (n1) at (0,3.5) {\( \text{n} \)};
				\vertex[right=4cm of n1] (p1) {\( \text{p} \)};
				\vertex (n2) at (0,0) {\( \text{n} \)};
				\vertex[right=4cm of n2] (p2) {\( \text{p} \)};
				
				\vertex at ($(n1)!0.4!(p1)$) (mid1) {};
				\vertex at ($(n2)!0.4!(p2)$) (mid2) {};
				
				\vertex[below right=1.5cm of mid1, dot, minimum size=0pt, inner sep=0pt] (e1) {};
				\vertex[above right=1.5cm of mid2, dot, minimum size=0pt, inner sep=0pt] (e2) {};
				
				\vertex[above right=0.4cm and 1.3cm of e1] (electron1) {\( e^- \)};
				\vertex[below right=0.4cm and 1.3cm of e2] (electron2) {\( e^- \)};
				
				\vertex[minimum size=0pt, inner sep=0pt] at ($(e1)!0.5!(e2)$) (mass) {\( \times \)};
				
				\diagram* {
					(n1) -- [draw=none] (p1),
					(n2) -- [draw=none] (p2),
					
					(mid1) -- [boson, edge label'=\( W^- \)] (e1),
					(mid2) -- [boson, edge label=\( W^- \)] (e2),
					
					(e1) -- [fermion] (electron1),
					(e2) -- [fermion] (electron2),
					
					(e1) -- [anti fermion, edge label'=\( \bar{\nu}_e \)] (mass),
					(e2) -- [anti fermion, edge label =\( \bar{\nu}_e \)] (mass),
				};
				
				\foreach \shift in {-0.1, 0, 0.1} {
					\draw[thick] ($(n1)+(0.2,\shift)$) -- ($(p1)+(-0.2,\shift)$);
					\draw[thick] ($(n2)+(0.2,\shift)$) -- ($(p2)+(-0.2,\shift)$);
				}
				
				\node[draw=none, fill=cyan!20, rounded corners, minimum width=1.7cm, minimum height=1cm] at (0,2.5) {\Large \( 0\nu\beta\beta \)};
			\end{feynman}
		\end{tikzpicture}
		\caption{}
		\label{0vbb}
	\end{subfigure}
	\end{minipage}
	}
	\caption{Tree-level Feynman diagram of two neutrinos double beta decay (a) and neutrinoless double beta decay in the hypothesis of light Majorana neutrino exchange (b)}
	\label{feynmand}
\end{figure}
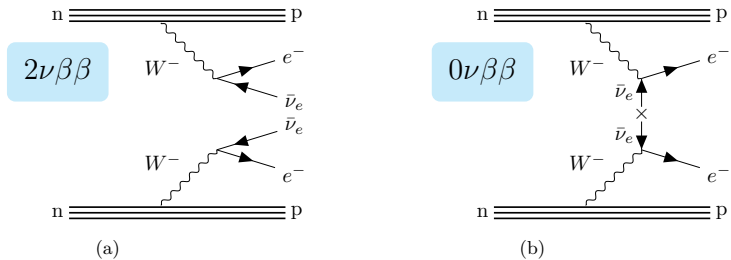

Two neutrinos double beta decay (\twovbb, \cref{2vbb}) is a rare process in which two nucleons in a nucleus simultaneously undergo a beta decay: two electrons and two neutrinos are produced and appear in the final state.
This process, despite being highly suppressed (second order weak process, $T_{1/2}^{2\nu} \approx 10^{18}-10^{21}$ yr), is allowed within the current formulation of the Standard Model (SM), as it entails lepton number conservation ($\Delta L = 0$), and has been experimentally observed for various even-even isotopes for which the single $\beta$ decay is energetically forbidden\tild\cite{Saakyan_2vbb_review}. 

\cref{0vbb} instead shows the Feynman diagram for the neutrinoless double beta decay (\novbb{}): in its final state only the two electrons appear, while the two neutrinos are missing.
This process, clearly entailing a lepton number violation of two units ($\Delta L = 2$), is not allowed within the current formulation of the SM, could take place only if neutrinos are Majorana fermions and has never been observed experimentally yet.

The half life of \novbb{} decay ($T_{1/2}^{0\nu}$) is related to the effective Majorana mass of the neutrinos $m_{\beta\beta}=\sum_i \big|U_{ei}^2m_i\big|$ by Eq.\tild(\ref{half_life}), in which $ G^{0\nu}(Q_{\beta\beta}, Z)$ is the phase space integral, $M^{0\nu}$ the nuclear matrix element and $m_e$ the mass of the electron.	
\begin{equation}
	[T_{1/2}^{0\nu}]^{-1} = G^{0\nu}(Q_{\beta\beta}, Z) |M^{0\nu}|^{2} \frac{|m_{\beta\beta}|^{2}}{m_e^2}
	\label{half_life}
\end{equation}
Considering that 
$ G^{0\nu}(Q_{\beta\beta}, Z)$ and $M^{0\nu}$ can be calculated, if $T_{1/2}^{0\nu}$ is measured experimentally it would be possible to extract $m_{\beta\beta}$ and therefore to obtain some information about the value of the sum of the neutrino mass eigenvalues.

But how can we look for \novbb{} experimentally and how do we distinguish it from the \twovbb{}?
The strategy is to measure the sum energy spectrum of the two emitted electrons.
Due to the different kinematics of \twovbb{} and \novbb{}, the contributions to the spectrum of the two different decays are clearly separated. 
Assuming the nuclear recoil energy is negligible,
\twovbb{} is a four body decay, therefore the sum of the energy of the two electrons has a continuous distribution with values between 0 and the Q value of the decay ($Q_{\beta\beta}$); \novbb{} instead is a two body decay, so the sum of the energy of the two electrons is exactly equal to $Q_{\beta\beta}$. The experimental signature of neutrinoless double beta decay is therefore a sharp peak at the endpoint of the $\beta\beta$ spectrum, where the \twovbb{} distribution would otherwise go to zero. 

Multiple $\beta\beta$ decaying isotopes can be exploited for the search of neutrinoless double beta decay, including $^{136}$Xe (EXO, KamLAND-Zen), $^{76}$Ge (GERDA, MAJORANA Demonstrator), $^{130}$Te (CUORE), $^{100}$Mo (CUPID-Mo, NEMO-3), $^{82}$Se (CUPID-0, NEMO-3). 
The LEGEND experiment uses $^{76}$Ge, looking for the decay $^{76}$Ge $\rightarrow$ $^{76}$Se + 2e$^-$ with $Q_{\beta\beta}$ =  2039.061(7) keV\tild\cite{qbb_value}.

\section{LEGEND(-200)}

The LEGEND project, building upon the expertise and results of the previous GERDA and MAJORANA Demonstrator experiments, aims at developing a two-phased experimental campaign. 
The first phase (current generation), LEGEND-200, foresees the installation of 200 kg of germanium; with a total exposure of 1 ton yr and a target background index (BI) of $2\cdot 10^{-4}$ counts/(keV kg yr), it is planned to reach 3$\sigma$ discovery sensitivity for a \novbb{} half life of about $10^{27}$ yr, corresponding to $m_{\beta\beta} \approx 33 - 89$ meV.
The second phase (next generation), LEGEND-1000, foresees the installation of 1000 kg of germanium; with a total exposure of 10 ton yr and a target BI of 10$^{-5}$ counts/(keV kg yr), it is planned to reach 3$\sigma$ discovery sensitivity for a \novbb{} half life beyond 10$^{28}$ yr, corresponding to $m_{\beta\beta} \approx 9 - 24$ meV.
\vspace*{-0.2cm}
\begin{figure}[H]
	\centering
	\includegraphics[width=0.4\textwidth]{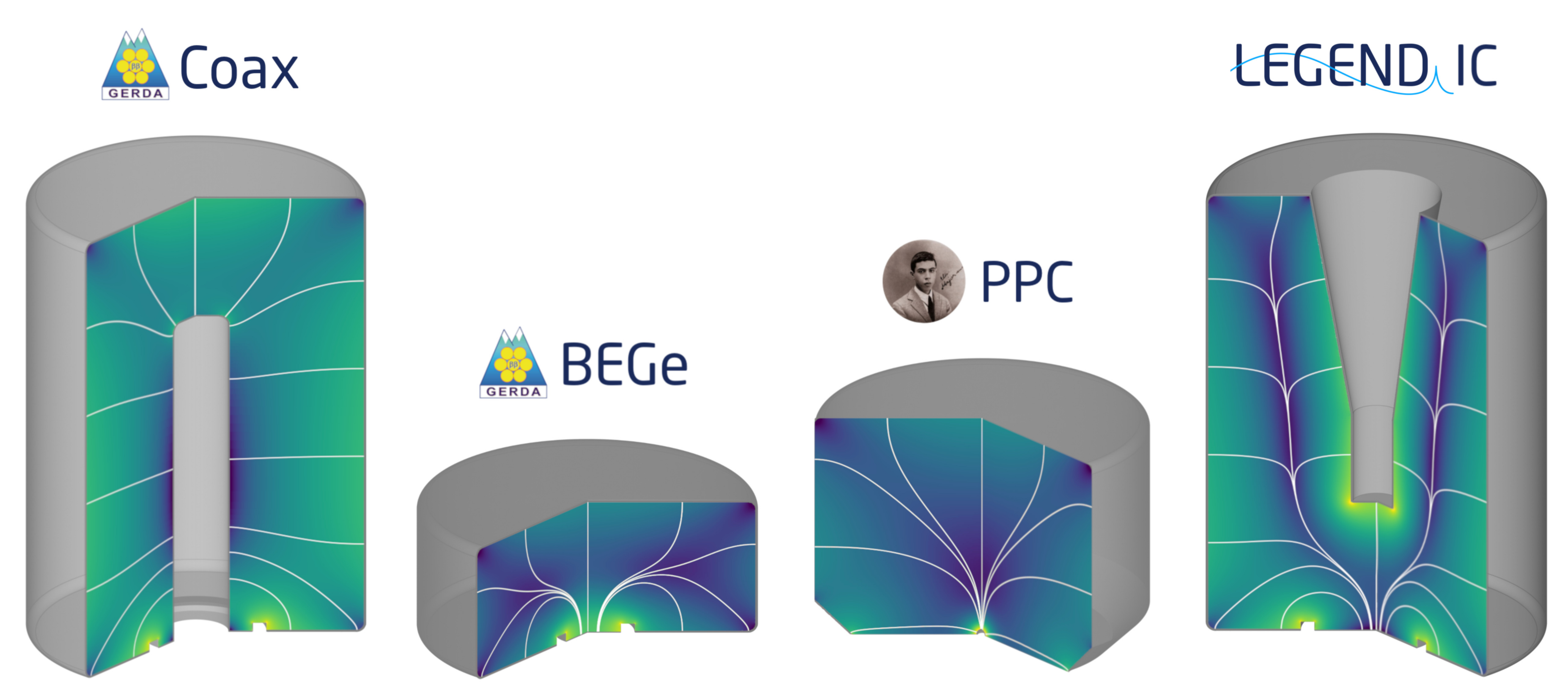}
	\caption{LEGEND-200 HPGe detector geometries}
	\label{legend-hpges}
	
\end{figure}
\vspace*{-0.2cm}
After a commissioning period with 60 kg of Ge, LEGEND-200 started taking physics data in March 2023 with a total Ge mass of 142.5 kg. 
LEGEND detectors are active High Purity Germanium (HPGe) detectors, enriched in $^{76}$Ge up to a fraction between 86\% and 92\% (depending on the detector type) 
from a natural abundance of $\sim$ 8\%\tild\cite{enr_ge}: their intrinsic background level is extremely low and their energy resolution is excellent, reaching a FWHM($Q_{\beta\beta}$) $\sim$ 0.1\%.
The employed germanium crystals have masses ranging between $\sim$ 0.5-4 kg and are operated as fully depleted diodes by applying a suitable high voltage: when an interaction takes place in the germanium, the released energy ionizes the atoms and the produced charge cloud drifts towards the electrodes, where it can be collected and amplified. The amplitude of such charge signal is proportional to the energy released in the interaction, which is the observable needed for the \novbb{} search.
Four types of germanium detectors are currently used in LEGEND-200 (Fig. \ref{legend-hpges}): 6 Coax and 28 BEGe from GERDA, 26 PPC from MAJORANA Demonstrator, and 41 Inverted Coaxial Point Contact (IC) detectors newly produced for LEGEND. 

BEGe, PPC and IC geometries are particularly convenient for LEGEND because the resulting electric field inside the detector is such that the shape of the produced signals is correlated to the topology of the energy deposition in the germanium. 
In \novbb{} events, the two electrons interact within $\sim$ 1 mm$^3$, creating a highly localized, single-site energy deposition; this is not the case for many background events, particularly gamma interactions, which produce multi-site depositions.
The combination of the detectors' sensitivity to different topologies and the unique topology signature of \novbb{} events enables therefore accurate background rejection using pulse shape discrimination (PSD) techniques.

\begin{figure}[H]
	\centering
	\vspace*{-0.2cm}
	\includegraphics[width=0.55\textwidth]{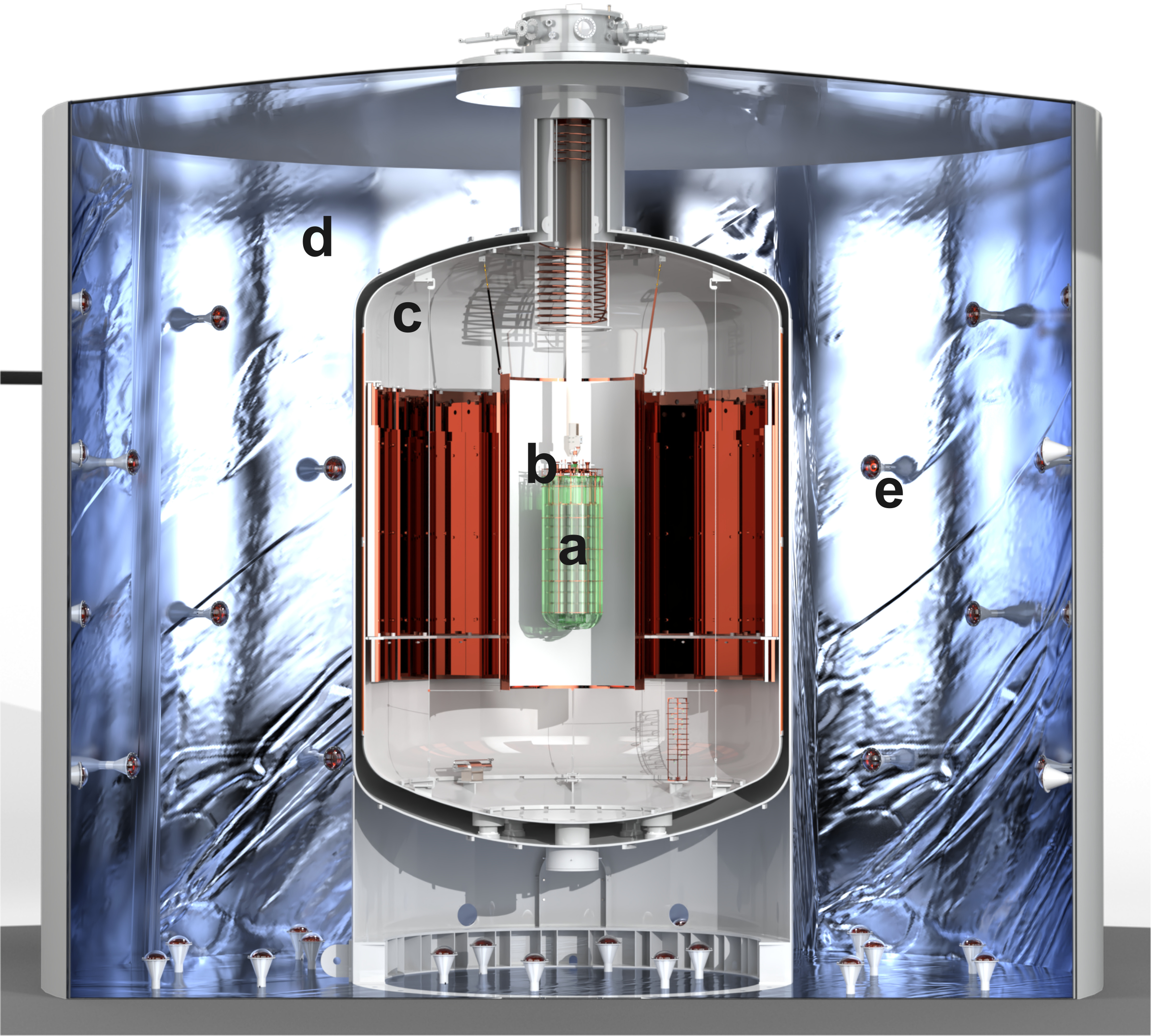}
	\caption{The LEGEND-200 experiment. Its main components are highlighted: a) Ge detectors array; b) WLS fibers barrel for LAr veto; c) LAr cryostat; d) Water tank; e) PMTs}
	\label{l200}
\end{figure}

The HPGe detectors array is the core of the LEGEND-200 experiment, represented in Fig.\tild\ref{l200}.
These detectors are operated bare in a 64 m$^{3}$ Liquid Argon (LAr) cryostat at $\sim$ 88 K.
LAr serves not only as a coolant and as a passive shield, but also as an active background veto, thanks to the instrumentation deployed in it: two barrels of wavelength-shifting (WLS) fibers are coupled at each end to silicon photomultipliers (SiPMs) to detect the scintillation light produced when an interaction takes place within the argon volume, allowing Ge-LAr anti-coincidence analysis. 
The LAr cryostat is placed in a 590 m$^3$ tank filled with ultra-pure water and instrumented with photomultiplier tubes (PMTs), which serves both as a passive shield and as Cherenkov detector for the few cosmic muons reaching the experiment despite its underground location: the 1400 m rock overburden above LNGS provides a shield of almost 3500 m water equivalent, which removes the hadronic component of cosmic ray showers and reduces the muon flux by almost six orders of magnitude with respect to the surface, to a rate of about 1.2 muons/(m$^2$ h)\tild\cite{muon_flux}.

\section{First results from LEGEND-200}

In approximately one year of data taking, LEGEND-200 collected $\sim$ 61 kg yr of data usable for the \novbb{} analysis. During this period, the performance of the experiment and the stability of the array have been thoroughly investigated.

\begin{figure}[H]
	\centering
	\begin{minipage}{.48\textwidth}
		\centering
		\includegraphics[width=\textwidth]{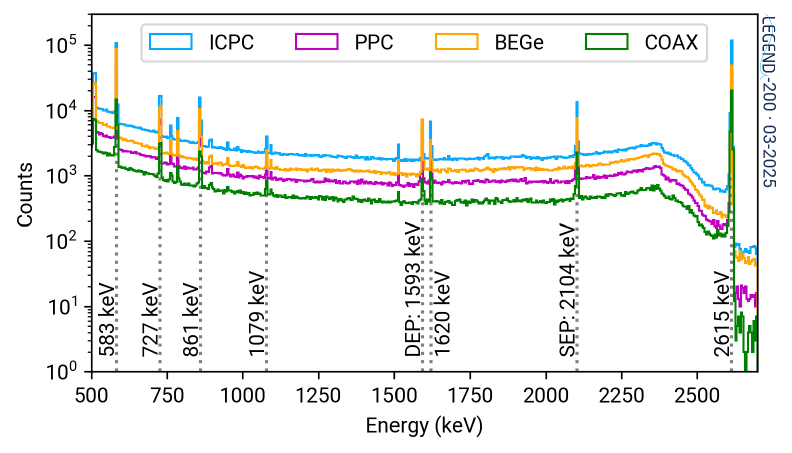}
		\caption{$^{228}$Th calibration spectrum}
		\label{cal_spectra}
	\end{minipage}%
	\hfill
	\begin{minipage}{.48\textwidth}
		\centering
		\includegraphics[width=\textwidth]{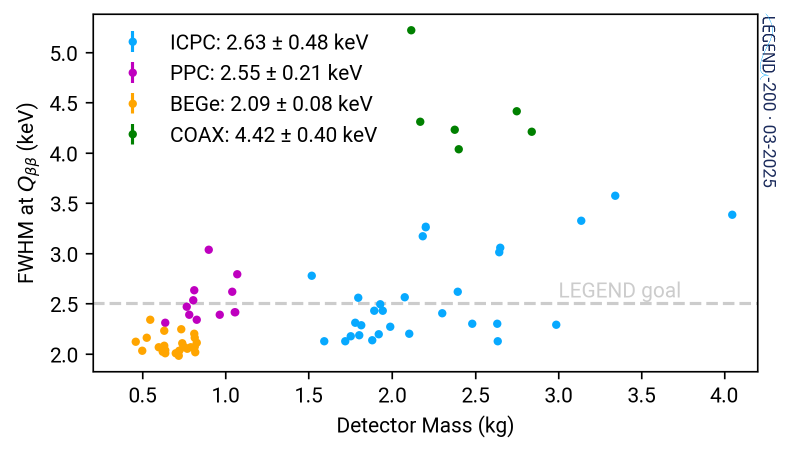}
		\caption{Energy resolution at $Q_{\beta\beta}$}
		\label{eres_qbb}
	\end{minipage}
\end{figure}

Physics runs were alternated to weekly calibrations performed with 13 $^{228}$Th sources of $\mathcal{O}$(1 kBq) activity each. Fig.\tild\ref{cal_spectra} shows the calibration spectra acquired by each detector type.
The FWHM of each peak is calculated to estimate each detector's energy resolution, as a function of the energy. This distribution is then fitted to extrapolate the detector's energy resolution at $Q_{\beta\beta}$. Fig.\tild\ref{eres_qbb} displays such energy resolutions as a function of the detector mass, color-coded by detector type: for most of the detectors the result is compatible with the LEGEND goal of FWHM($Q_{\beta\beta}$) $\leq$ 2.5 keV. A clear exception is COAX detectors, whose energy resolution was though known to be relatively poor from GERDA measurements\tild\cite{final_res_gerda}: this will not be an issue in LEGEND-1000, in which only ICPC are planned to be deployed.\\

\begin{figure}[H]
	\centering
	\includegraphics[width=0.95\textwidth]{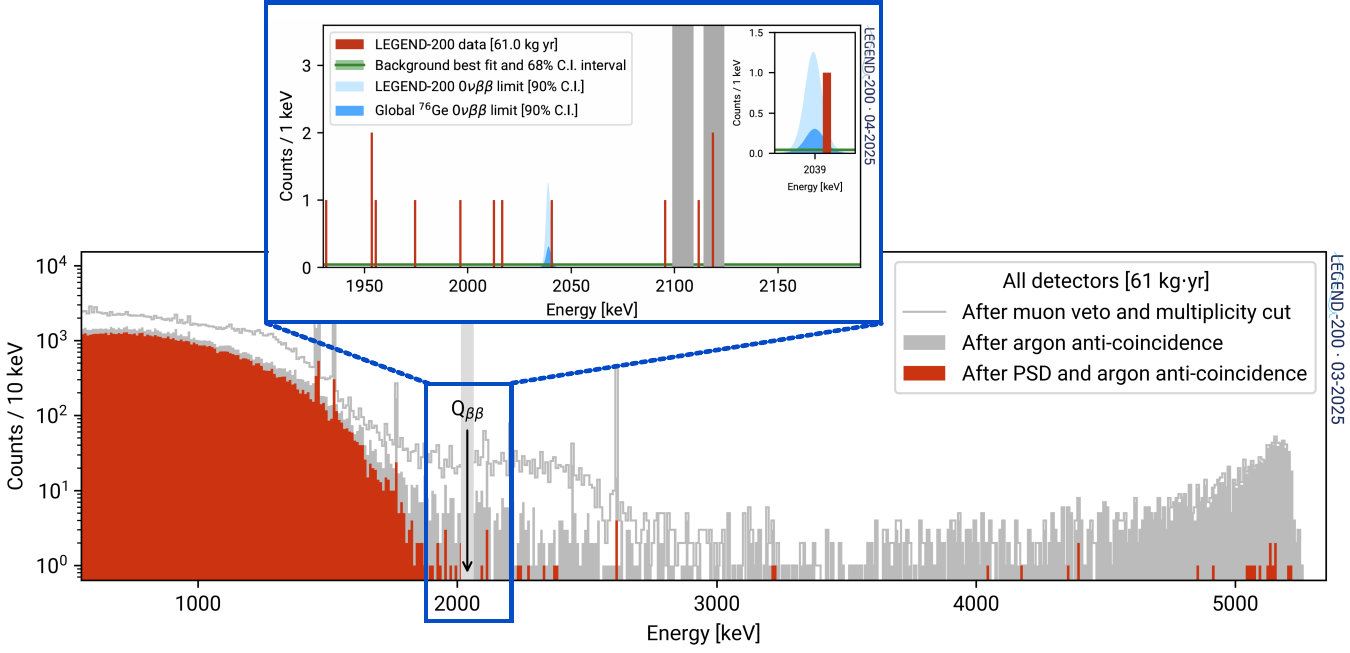}
	\caption{LEGEND-200 energy spectrum. The inset shows the unblinded analysis window.}
	\label{phy_spectrum}
	\vspace*{-0.2cm}
\end{figure}

The acquired physics energy spectrum is displayed in Fig.\tild\ref{phy_spectrum}: different colors show the result of the application of the different analysis cuts.
The spectrum displayed in white contains all the events that pass the muon veto and the Ge detectors anti-coincidence cut - defined \textit{multiplicity cut} -, both having efficiency $\epsilon_{muon}(Q_{\beta\beta}),\epsilon_{mult}(Q_{\beta\beta}) > 99.9$\%. 
This spectrum exhibits all the expected features; however, when compared to background predictions based on the available radioassay data, it indicates that the background level is higher than expected, also in the region around $Q_{\beta\beta}$. 
The background model fit of this spectrum suggests that this excess could be imputable to the $^{228}$Th chain: a thorough screening and cleaning campaign has been conducted to identify and eliminate the source of this contamination; confirmation of the expected background reduction will come with the analysis of the data acquired after the redeployment of the array in May 2025.
The spectrum of the events passing the LAr veto is displayed in grey: this cut is particularly effective at suppressing Compton event, and its efficiency reaches $\epsilon_{LAr}(Q_{\beta\beta}) \sim 93$\%.
Finally the spectrum of the events passing also the PSD cut is displayed in red: here the \twovbb{} distribution is clearly recognizable in the lower energy region, while only few residual background events appear at higher energies.
The efficiency of the PSD cut reaches $\epsilon_{PSD}(Q_{\beta\beta}) \sim 76-85$\% depending on the detector type.

All the data selection routines are optimized and applied in a strictly blind scheme in the region $Q_{\beta\beta} \pm 25$ keV, displayed in light gray.
The energy range considered for the statistical analysis aimed at the calculation of the background index (BI) and of the half life limit is [1930-2190] keV with the exclusion of two known gamma lines, $[2104\pm5]$ keV from $^{208}$Tl and $[2119\pm5]$ keV from $^{214}$Bi.

Two separate unblindings have been performed for two different subsets of the acquired data, which are expected to have different background levels.
The so called \texttt{golden dataset} consists of 48.3 kg yr of exposure coming from BEGe, PPC and ICPC produced by Mirion, while the so called \texttt{silver dataset} consists of 12.7 kg yr of exposure coming from COAX and ICPC produced by ORTEC. 
A total of 11 events survive all the cuts in the analysis window (inset of Fig.\tild\ref{phy_spectrum}), 7 from the \texttt{golden dataset} and 4 from the \texttt{silver dataset}.

The frequentist analysis finds no evidence of a \novbb{} signal and sets a lower limit on its half life to $T_{1/2}^{0\nu} > 0.5 \cdot 10^{26}$ yr at 90\% confidence level (CL), with a median exclusion sensitivity of $1.0\cdot10^{26}$ yr.
The achieved background index results to be $0.5_{-0.2}^{+0.3} \cdot 10^{-3}$ counts/(keV kg yr) in the \texttt{golden dataset} and $1.3_{-0.5}^{+0.8} \cdot 10^{-3}$ counts/(keV kg yr) in the \texttt{silver dataset}.  
A combined GERDA (127.2 kg yr) + MAJORANA Demonstrator (64.5 kg yr) + LEGEND-200 (61 kg yr) fit is finally performed: also in this case no evidence of a \novbb{} signal is found and a lower limit to its half life is set instead to $T_{1/2}^{0\nu} > 1.9 \cdot 10^{26}$ yr at 90\% CL, with a median exclusion sensitivity of $2.8\cdot 10^{26}$ yr. The corresponding Bayesian analysis provides compatible results.
Constraints on the effective Majorana mass are calculated in the frequentist framework using a range of nuclear matrix elements from phenomenological calculations: the obtained upper limits range between $m_{\beta\beta} <$  75–200 meV \cite{prl25}.

\section{Conclusions}

The LEGEND-200 experiment successfully completed the first year of physics data taking collecting the first 61 kg yr of exposure. The background level appearing in those data is slightly higher than expected from radioassays prediction: for this reason an extensive screening and cleaning campaign was undertaken, whose outcome is expected to be clear by the end of summer 2025.
The background excess is anyway successfully mitigated by the implemented analysis cuts, in particular by the combination of LAr veto and PSD cut.
The frequentist analysis of the acquired data finds no evidence of \novbb{} signal and sets a limit of $T_{1/2}^{0\nu} > 0.5 \cdot 10^{26}$ yr at 90\% CL with a BI of $0.5_{-0.2}^{+0.3} \cdot 10^{-3}$ counts/(keV kg yr) in the \texttt{golden dataset} and $1.3_{-0.5}^{+0.8} \cdot 10^{-3}$ counts/(keV kg yr) in the \texttt{silver dataset}.
The combined GERDA, MAJORANA Demonstrator and LEGEND fit sets the $^{76}$Ge \novbb{} half life limit to $T_{1/2}^{0\nu} > 1.9 \cdot 10^{26}$ yr at 90\% CL.\\

\noindent
\textbf{Acknowledgements} This work is supported by the U.S. DOE, and the NSF, the LANL, ORNL and LBNL LDRD programs; the European ERC and Horizon programs; the German DFG, BMBF, and MPG; the Italian INFN; the Polish NCN and MNiSW; the Czech MEYS; the Slovak RDA; the Swiss SNF; the UK STFC; the Canadian NSERC and CFI; the LNGS and SURF facilities.

\end{document}